\documentclass{PoS}
\title{Fluctuations and correlations in nucleus-nucleus collisions
observed by NA61/SHINE at CERN SPS energies}

\ShortTitle{Fluctuations and correlations at NA61/SHINE}

\author{\speaker{Maja Ma\'{c}kowiak-Paw{\l}owska} for the NA61/SHINE Collaboration\\
        Faculty of Physics, Warsaw University of Technology\\
        E-mail: \email{maja.pawlowska@pw.edu.pl}}

\abstract{The NA61/SHINE strong interaction programme aims to explore the phase diagram of the strongly interacting matter. The main physics goals are the study of the onset of deconfinement and the search for the critical point of the strongly interacting matter. These goals are pursued by performing a scan in beam momentum (13$A$ -- 158$A$ GeV/c) and size of colliding system (p+p, p+Pb, Be+Be, Ar+Sc, Xe+La, Pb+Pb).

This contribution presents new results on system size and energy dependence of multiplicity and net-charge fluctuations measured with higher-order moments of these distributions. Also, new results on the measurement of Bose-Einstein correlations in nucleus-nucleus collisions based on Levy sources are discussed as well as news on anisotropic flow in Pb+Pb interactions at 13$A$ GeV/c.}

\FullConference{the International conference on Critical Point and Onset of Deconfinement\\
                15-19.03.2021\\
                Online}

\begin{document}
\section{Introduction}
NA61/SHINE~\cite{Antoniou:2006mh,Abgrall:2014xwa} at the CERN Super Proton Synchrotron (SPS) is a fixed-target experiment pursuing a rich physics program including measurements for strong interactions, neutrino, and cosmic ray physics.

\begin{figure}
			\centering	
			\includegraphics[width=0.45\textwidth]{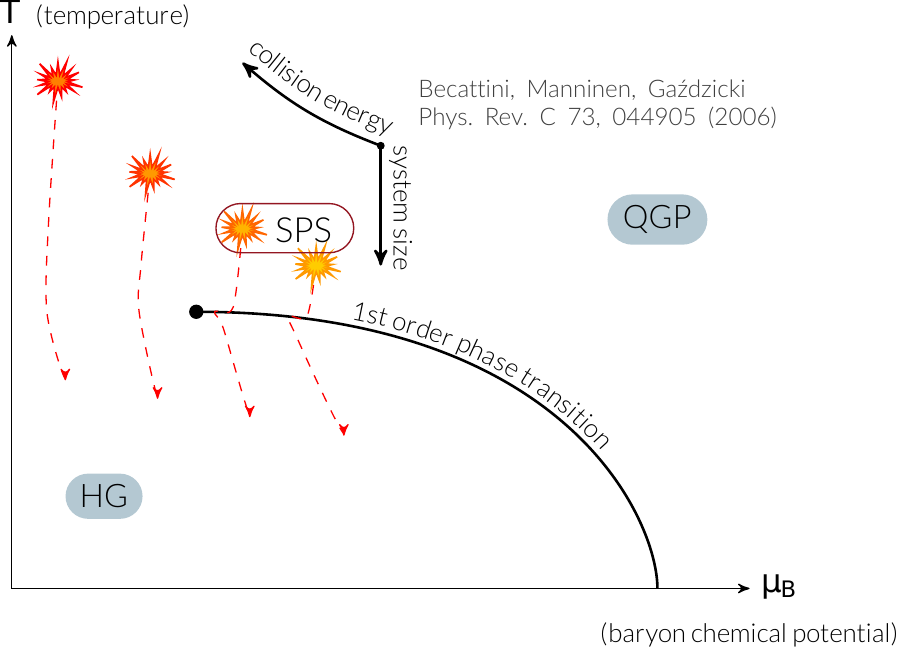}
			\includegraphics[width=0.45\textwidth]{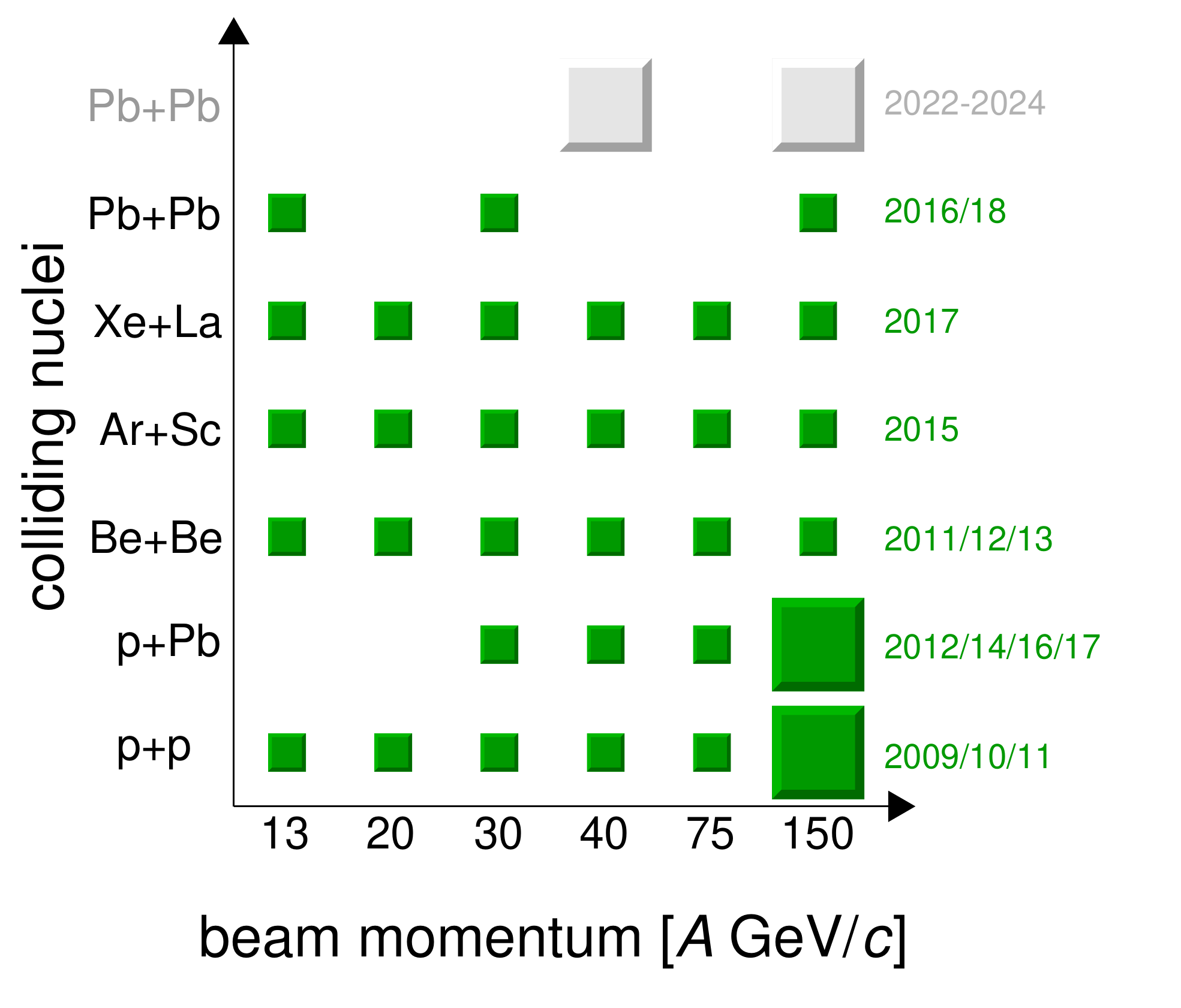}			
			\vspace{-0.15in}
			\caption{NA61/SHINE system size and energy scan}
			\label{fig:program}
	\end{figure}

The strong interactions program focuses on search for the critical point (CP) and study of the onset of deconfinement (OD) of strongly interacting matter. NA61/SHINE is the first experiment to perform a two-dimensional scan, in beam momentum (13$A$ -- 150/158$A$ GeV/$c$)  and size  of colliding system (p+p, p+Pb, Be+Be, Ar+Sc, Xe+La, Pb+Pb). It aims to explore the phase diagram of the strongly interacting matter. The illustration of the scan as well as list of gathered and planed system/energies are shown in Fig.~\ref{fig:program}.

This contribution discusses new results on fluctuation and correlations measured in p+p, Be+Be, Ar+Sc 	and Pb+Pb interactions.
	
\section{Study of the onset of deconfinement}
Spatial asymmetry of the initial energy density in the overlapping region of the colliding relativistic nuclei is converted, via interactions between produced particles, to the asymmetry of momentum distribution of particles in the final state. The resulting asymmetry carries information about the transport properties of the QCD matter created during the collision. Asymmetry is usually quantified with $v_n$ coefficients in a Fourier decomposition of the azimuthal distribution of produced particles relative to the reaction plane. The NA61/SHINE has an unique way to estimate the reaction plane with the Projectile Spectator Detector (for details see Refs.~\cite{Golosov:2019sdu,EKashirin}). 

\begin{figure}
			\centering	
			\includegraphics[width=0.32\textwidth]{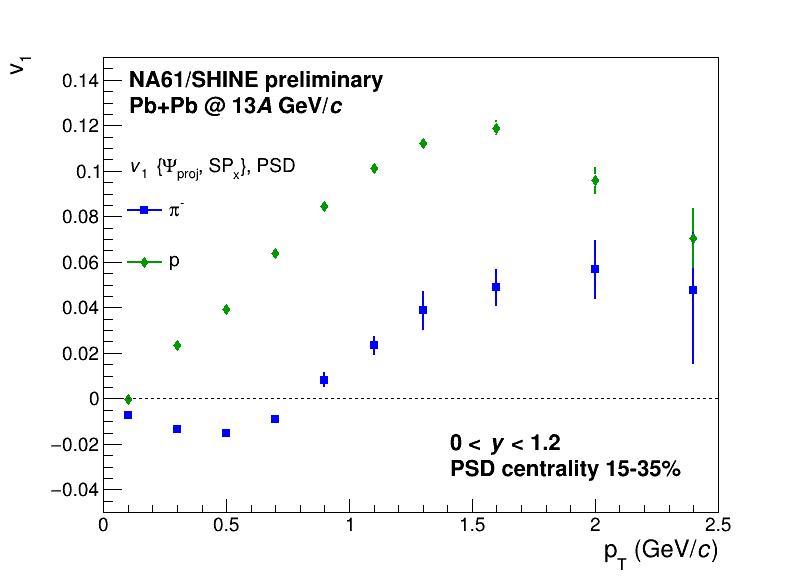}
			\includegraphics[width=0.305\textwidth]{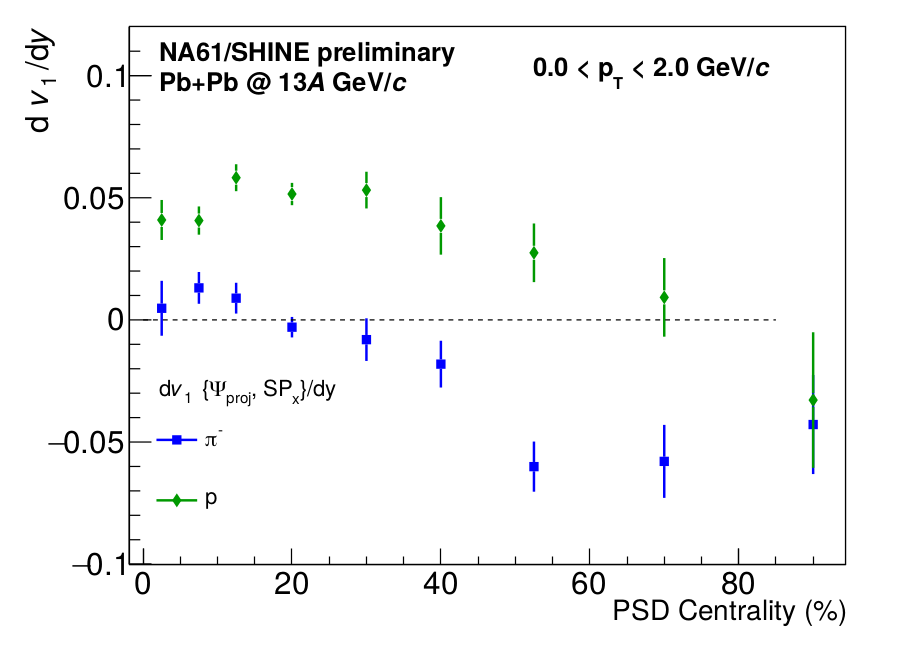}
			\includegraphics[width=0.32\textwidth]{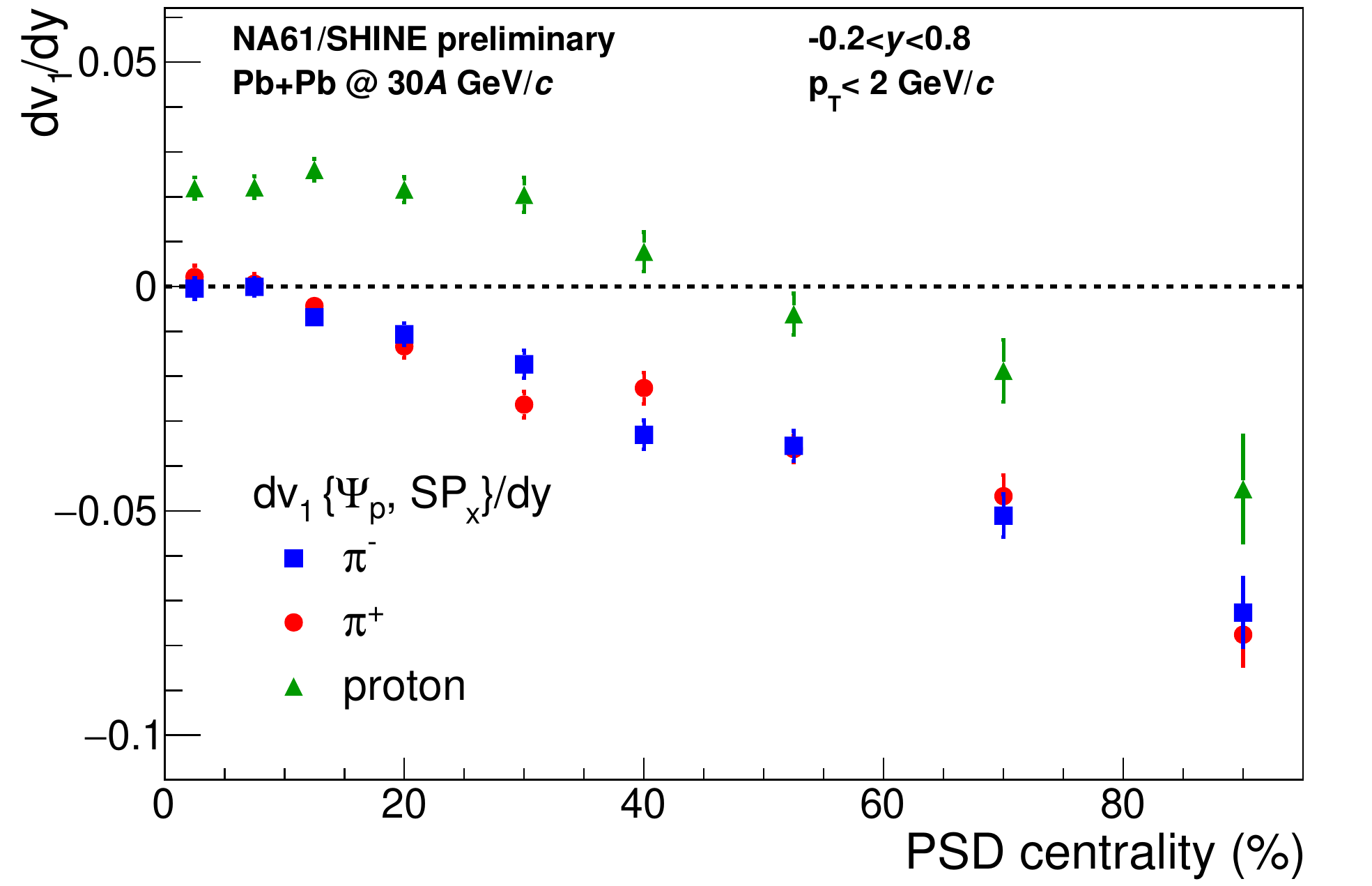}	
			\vspace{-0.15in}		
			\caption{Negatively charged pion and proton directed flow $v_{1}(p_{T})$ and $dv_{1}/dy$ for different centrality classes in Pb+Pb collisions.}
			\label{fig:flow}
	\end{figure}

The energy dependence of flow coefficients is of particular importance.  At the energies of SPS and Beam Energy Scan Program at RHIC it is expected that the slope of proton directed flow at mid-rapidity, $dv_1/dy$, changes its sign~\cite{STARv,STARv2,Wu:2018qih}. Directed flow of $\pi^{-}$ and $p$ as well as $dv_1/dy$ (centrality dependence) for Pb+Pb collisions at 13$A$ and 30$A$ GeV/$c$ is presented in Fig.~\ref{fig:flow}. Shapes of $v_{1}(p_{T})$ for protons and negatively charged pions (Fig.~\ref{fig:flow}, left) are different. $v_1(p_T)$ of protons is positive in the entire $p_T$ range. Directed flow of negatively charged pions starts with negative values and than changes sign (Fig.~\ref{fig:flow}, center). There is also a clear difference of the $v_{1}$ slope between 13$A$ and 30$A$ GeV/$c$ (Fig.~\ref{fig:flow}, center and right).
\section{Search for the critical point}

\begin{figure}
			\centering	
			\includegraphics[width=0.32\textwidth]{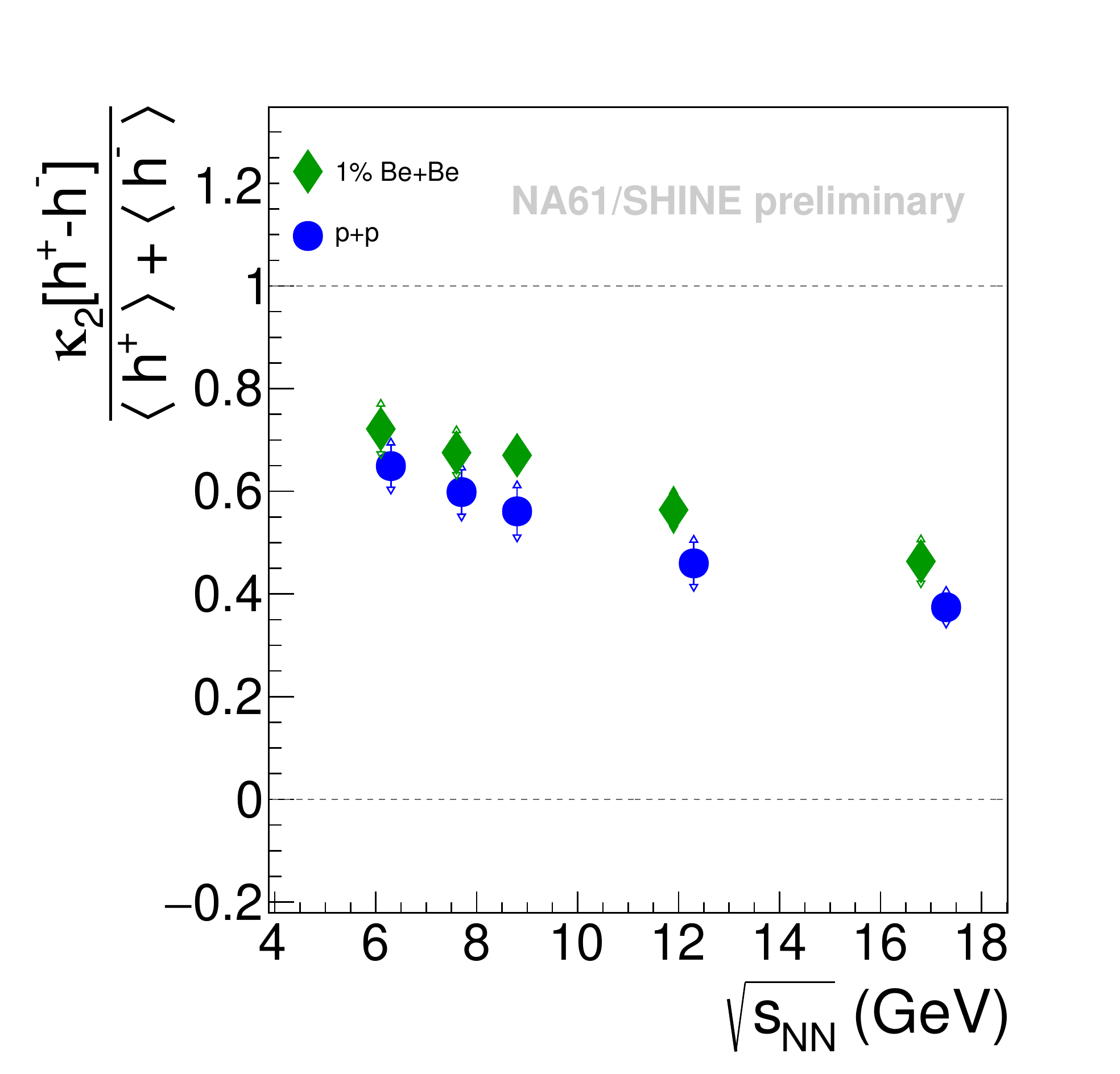}
			\includegraphics[width=0.32\textwidth]{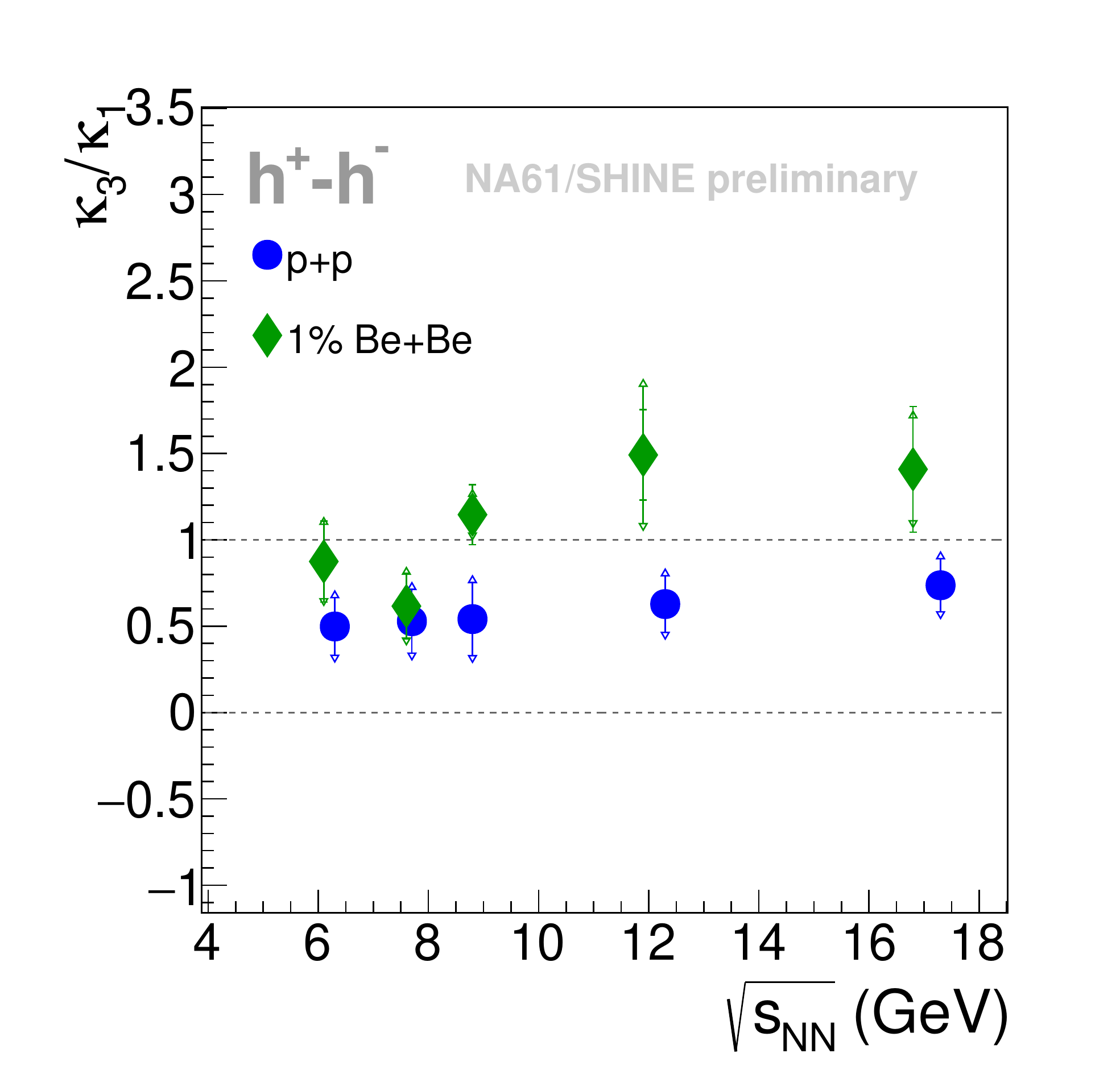}
			\includegraphics[width=0.32\textwidth]{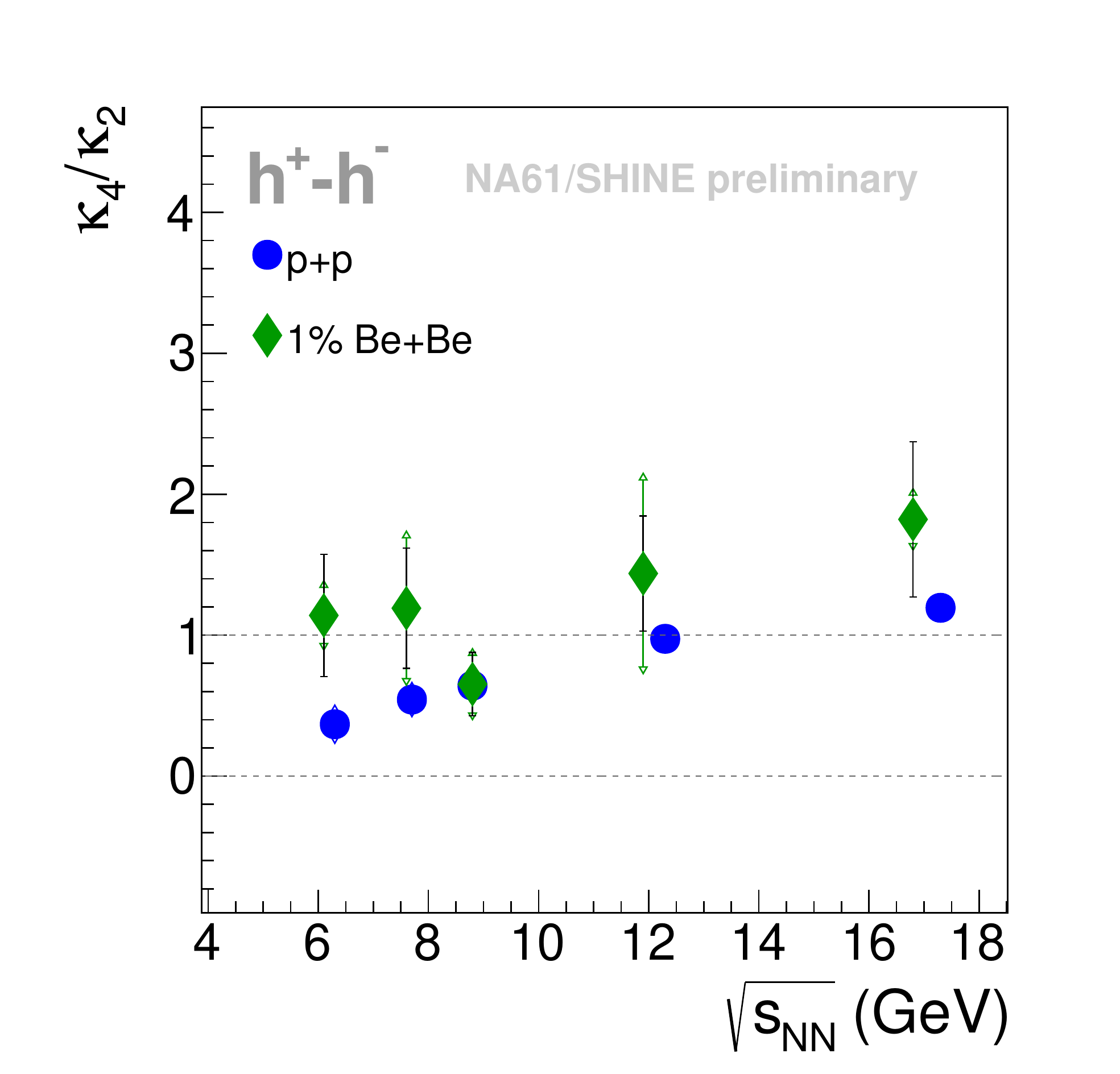}
			\vspace{-0.15in}
			\caption{System size and energy dependence of $\kappa_{2}/\kappa_{1}[h^{+}-h^{-}]$, $\kappa_{3}/\kappa_{1}[h^{+}-h^{-}]$ and $\kappa_{4}/\kappa_{2}[h^{+}-h^{-}]$. Statistical uncertainty was obtained with the bootstrap method and it is indicated as a dashed black bar. Systematic uncertainty/bias: p+p - corrected data with estimate on systematic uncertainty; Be+Be - uncorrected data with estimate of systematic bias. Systematic uncertainty/bias is indicated with a green bar.}
			\label{fig:net}
	\end{figure}
	\begin{figure}
			\centering	
			\includegraphics[width=0.32\textwidth]{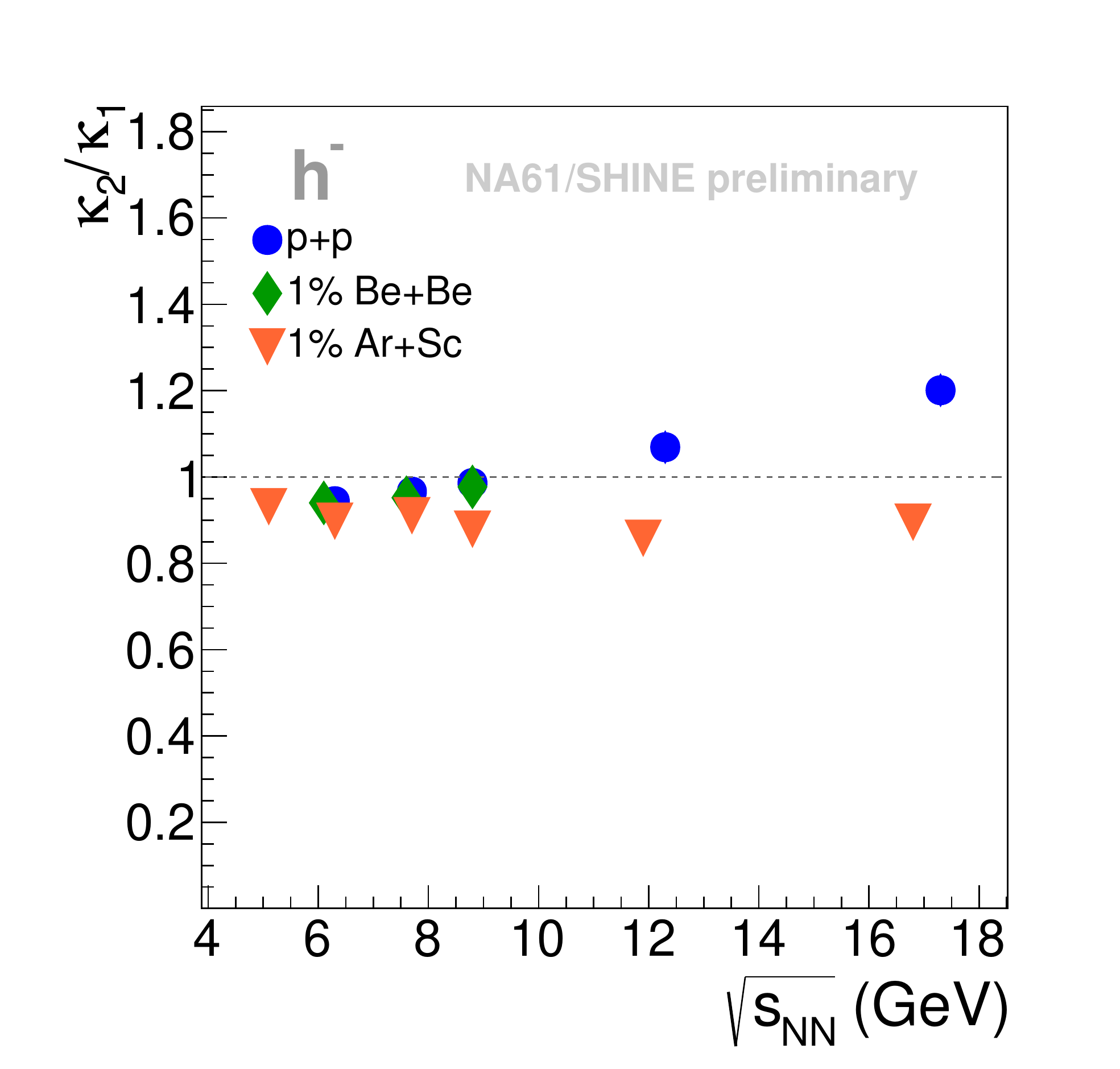}
			\includegraphics[width=0.32\textwidth]{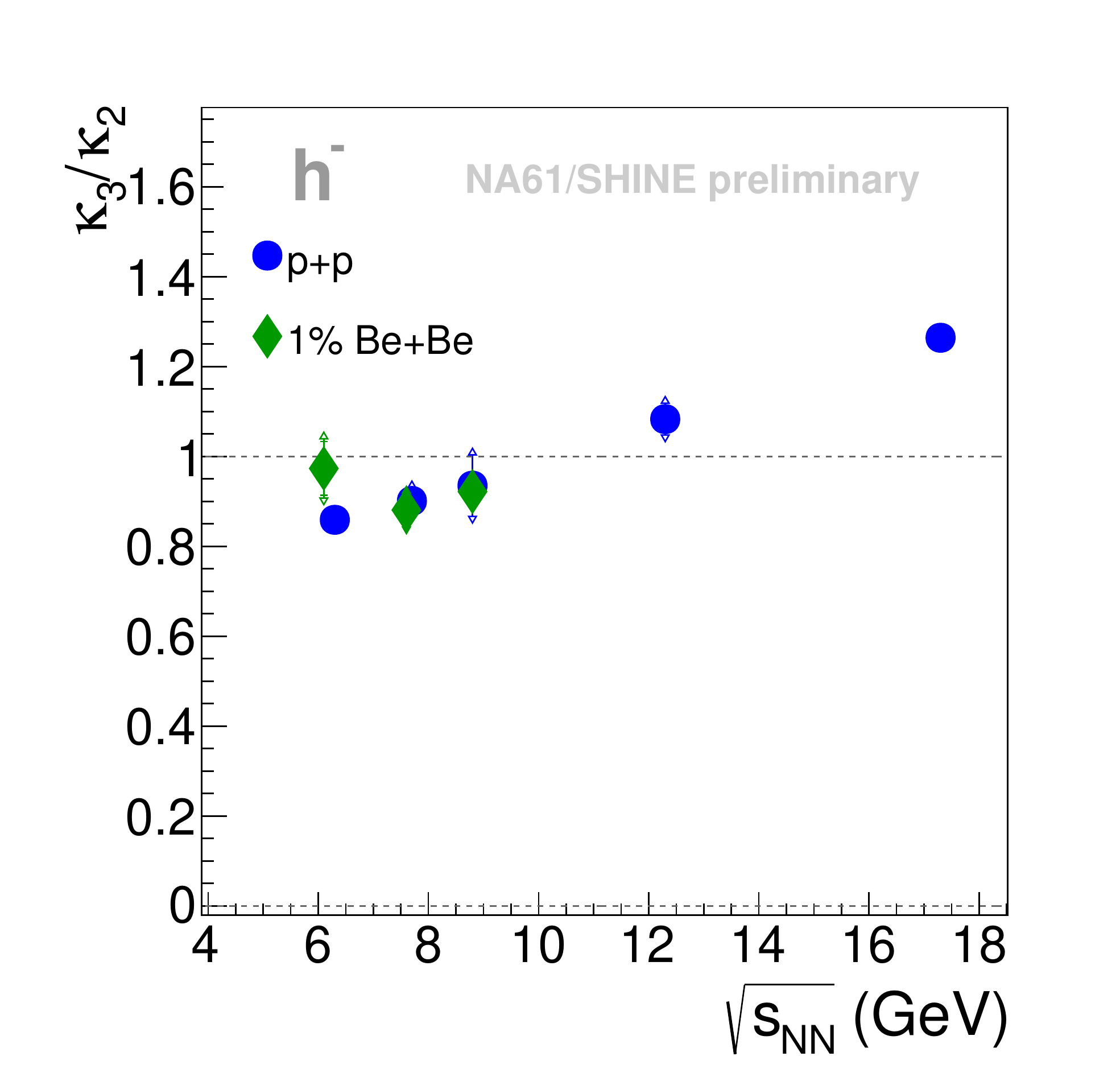}
			\includegraphics[width=0.32\textwidth]{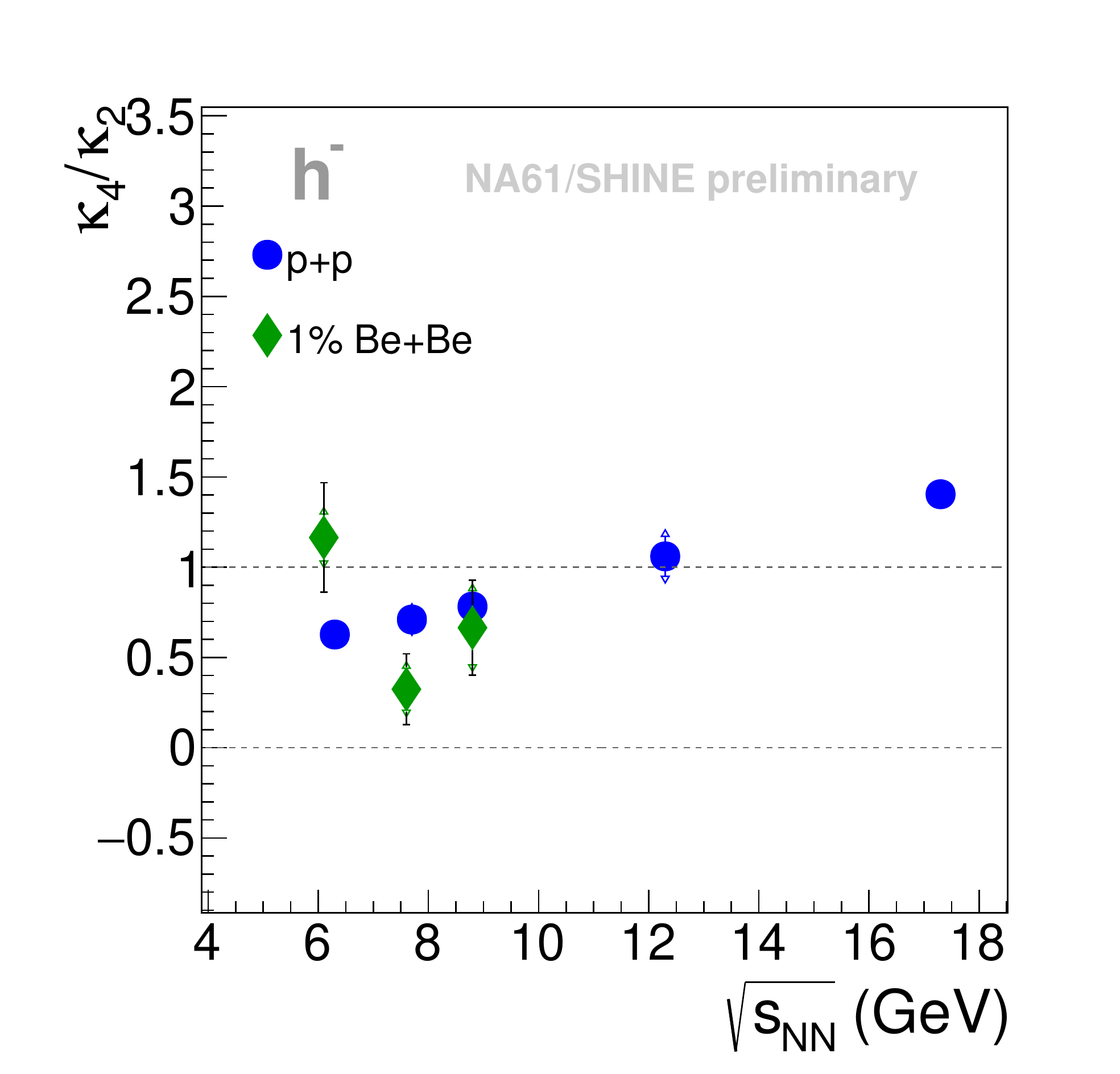}
			\vspace{-0.15in}
			\caption{System size and energy dependence of $\kappa_{2}/\kappa_{1}[h^{-}]$, $\kappa_{3}/\kappa_{2}[h^{-}]$ and $\kappa_{4}/\kappa_{2}[h^{-}]$. Statistical uncertainty was obtained with the bootstrap method and it is indicated as a dashed black bar. Systematic uncertainty/bias: p+p - corrected data with estimate on systematic uncertainty; Be+Be - uncorrected data with estimate of systematic bias. Systematic uncertainty/bias is indicated with a green bar.}
			\label{fig:neg}
	\end{figure}
	
	\begin{figure}
			\centering	
			\includegraphics[width=0.31\textwidth]{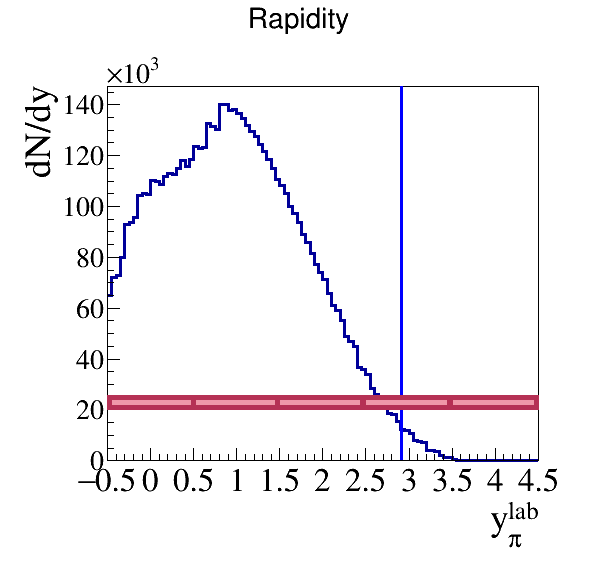}
			\includegraphics[width=0.31\textwidth]{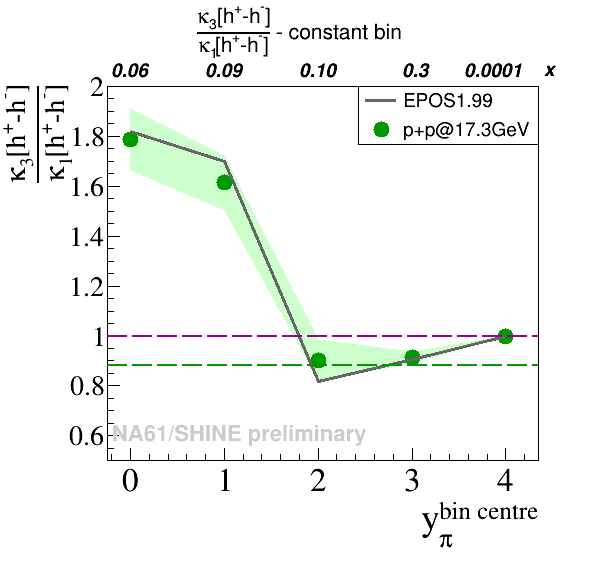}
			\includegraphics[width=0.31\textwidth]{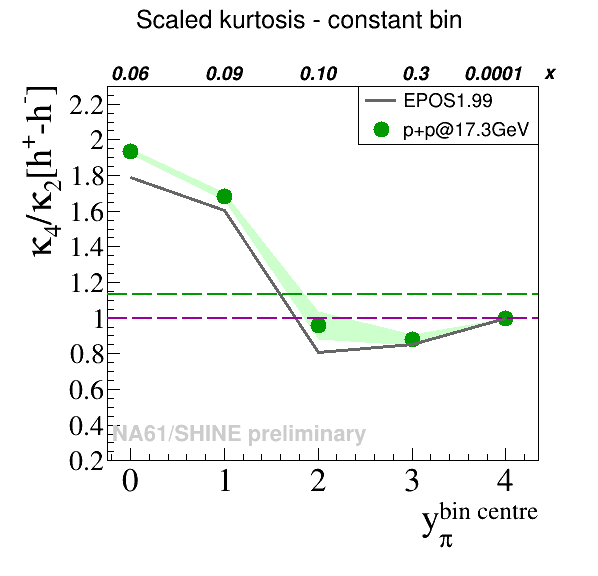}\\
			\includegraphics[width=0.31\textwidth]{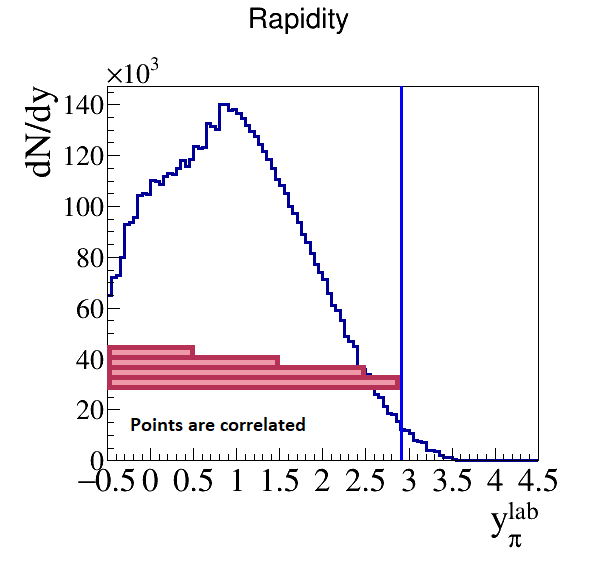}
			\includegraphics[width=0.31\textwidth]{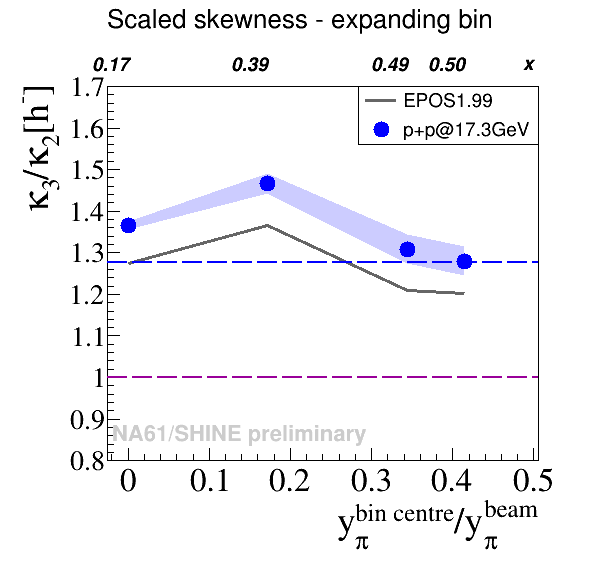}
			\includegraphics[width=0.31\textwidth]{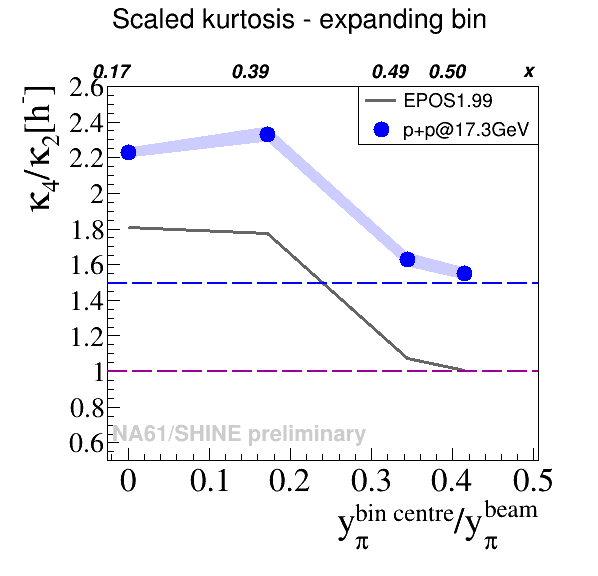}
			\vspace{-0.15in}
			\caption{Top: Bin selection and rapidity dependence of $\kappa_{3}/\kappa_{2}[h^{+}-h^{-}]$ and $\kappa_{4}/\kappa_{2}[h^{+}-h^{-}]$. Bottom: Bin selection and rapidity dependence of $\kappa_{3}/\kappa_{2}[h^{-}]$ and $\kappa_{4}/\kappa_{2}[h^{-}]$. Statistical uncertainty within the markers. Systematic uncertainty indicated with the color band.}
			\label{fig:yneg}
	\end{figure}	
	The expected signal of a critical point (CP) is a non-monotonic dependence of various fluctuation/correlation measures in NA61/SHINE energy -- system size scan. Special interest is devoted to fluctuations of conserved charges (electric, strangeness or baryon number)~\cite{Stephanov_overview, Asakawa:2015ybt}.

In order to compare fluctuations in systems of different sizes, one should use intensive quantities, i.e. quantities insensitive to system volume. Such quantities are constructed by division of cumulants $\kappa_{i}$ of the measured distribution (up to fourth order), where $i$ is the order of the cumulant. For second, third and fourth order cumulants intensive quantities are defined as: $\kappa_{2}/\kappa_{1}$, $\kappa_{3}/\kappa_{2}$ and $\kappa_{4}/\kappa_{2}$. Their reference values for multiplicity fluctuations are 0 (no fluctuations) and 1 (independent particle production). In case of net-charge, ratios are redefined to $\kappa_{2}/\kappa_{1}[h^{+}-h^{-}]$, $\kappa_{3}/\kappa_{1}[h^{+}-h^{-}]$ and $\kappa_{4}/\kappa_{2}[h^{+}-h^{-}]$ in order to keep the same reference. 

Figures~\ref{fig:net} and~\ref{fig:neg} show the system size and energy dependence of second, third and fourth order cumulant ratio (or its intensive equivalent) of net-electric charge and negatively charged hadron multiplicity in p+p, Be+Be and Ar+Sc interactions. So far, there is no clear difference between systems of different sizes except in case of $\kappa_{2}/\kappa_{1}[h^{-}]$ of p+p and Ar+Sc interactions. More detailed studies are needed.

Establishing a baseline is also an important part of the CP search. Preliminary results on rapidity dependence on net-charge and multiplicity fluctuations in p+p interactions at $\sqrt{s_{NN}}=17.3$~GeV are presented in Fig.~\ref{fig:yneg} top (constant bins) and bottom (widening bins). Results are compared with EPOS1.99~\cite{Pierog:2009zt,EPOSWeb}. It reproduces net-charge rapidity dependence but it underestimates signal in case of $h^{-}$.
\begin{figure}
			\centering	
			\includegraphics[width=0.52\textwidth]{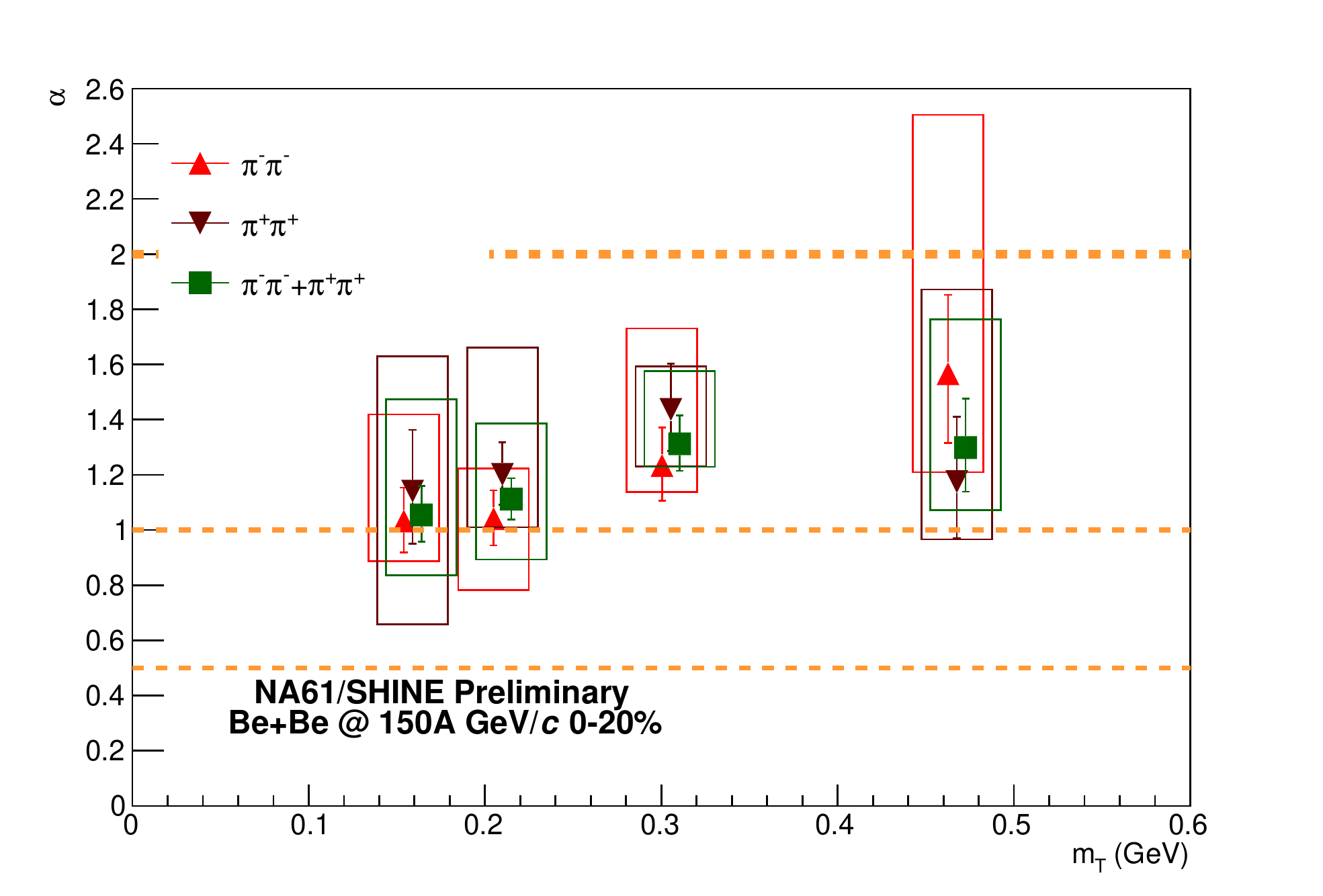}
			\vspace{-0.15in}
			\caption{Results on $\alpha$ parameter in Be+Be collisions at 150$A$ GeV/$c$. Statistical and systematic uncertainties are indicated with a bar and a box, respectively.}
			\label{fig:hbt}
	\end{figure}

At the CP, the spatial correlation function becomes a power-law $\sim r^{-(d-2+\eta)}$, where $d$ represents the number of  dimensions. 
One can predict a critical exponent $\eta$ (related to spatial correlations) for the QCD universality class, which is the 3D-Ising model for QCD~\cite{Stephanov:1998dy, Halasz:1998qr}. The predicted value of $\eta$ at the CP is 0.03631~\cite{El-Showk:2014dwa}. For the random field 3D Ising $\eta=0.50\pm0.05$~\cite{Rieger:1992th}. 
In the HBT analysis the momentum correlation function $C(q)$ of produced particles is directly related to the normalized source distribution $S(r)$ 
via $C(q) = 1 + (|\tilde{S}|)^2$, where $\tilde{S}$ is the Fourier transformation of $S(r)$. 
The data analysis was done by using a Levy distributed source function~\cite{Porfy:2019dtc}.  
Since, it leads to the same power-law tails, the Levy exponent $\alpha$ was assumed to be identical to the spatial correlation exponent $\eta$~\cite{Csorgo:2003uv}.  
In the vicinity of the critical point, $\alpha$ values around 0.5
may be expected and this can be measured by investigating the Bose-Einstein correlation function $C(q)=1+\lambda e^{-(qR)\alpha}$. 
Figure~\ref{fig:hbt} shows measured values of $\alpha$ parameter for pion pairs ($\pi^{-}\pi^{-}$, $\pi^+\pi^+$ and $\pi^{-}\pi^{-}+\pi^+\pi^+$) in 20$\%$ most central Be+Be collisions at 150$A$ GeV/$c$. 

All measured combinations (see Fig.~\ref{fig:hbt}) indicate $1<\alpha<2$ which is far from the CP value. In addition to the CP, $\alpha$ values lower than 2 can be caused by anomalous diffusion, QCD fractal structured jet fragmentation, and also to some extend by the averaging over broad event class (e.g. centrality)~\cite{Csanad:2007fr, Csorgo:2004sr, Csorgo:2005it, Cimerman:2019hva}.

\section{Summary}
In this contribution new results on fluctuations and correlations from the NA61/SHINE experiment were discussed. Directed flow $v_1$ is measured relative to the spectator plane in Pb+Pb collisions at 13$A$ GeV/$c$ for $p$ and $\pi^{-}$ and compared to results in Pb+Pb at 30$A$ GeV/$c$. We observe strong centrality dependence of directed flow for negatively charged pions. $v_1(p_T)$ of $\pi^{-}$ changes sign at $p_T \sim$ 1 GeV/$c$. The slope of $v_1(y)$ for pions (negative slope) have different sign compared to the one observed for protons (positive slope) in the specific centrality range. Multiplicity and net-charge fluctuations measured by higher order cumulant ratios were compared for different system sizes and energies. Net-charge results are comparable between systems but there is a difference in $\kappa_{2}/\kappa_{1}[h^{-}]$ for p+p and Ar+Sc collisions. The reference measurements on net-charge and multiplicity fluctuations in p+p as a function of rapidity were also reported. EPOS1.99 describes the data except the multiplicity fluctuations, where the description is qualitative only.  Value of $\alpha$ parameter of HBT analysis is between 1 and 2 in Be+Be at 150$A$ GeV/$c$ interactions possibly due to anomalous diffusion or other phenomena. In general, presented experimental results show no indications of the critical point. In order to qualitatively measure the CP signal, the background phenomena as well as remaining analysis should be studied in details.

\textbf{Acknowledgments:} This work was supported by WUT-IDUB and the National Science Centre, Poland under grant no. 2016/21/D/ST2/01983.

\end{document}